\documentclass[12pt]{iopart}

\usepackage{amstext}
\usepackage{epsfig} 
\usepackage{graphicx}

\begin{document}

\title[]{The Flavours of the Quark-Gluon Plasma}

\author{Berndt M\"uller}

\address{Department of Physics, Duke University, Durham, NC 27708, USA}
\ead{muller@phy.duke.edu}

\begin{abstract}
Quarks of other flavours than up and down, i.e.\ $s$, $c$, and $b$ quarks, have been long recognized as effective probes of the structure of hot QCD matter. In this talk, I review some of the motivations for their investigation and discuss the salient results obtained so far, with a focus on the results from the Relativistic Heavy Ion Collider (RHIC). Some ideas for future studies are also mentioned.
\end{abstract}



\section{The QGP is a ``Strange'' State of Matter}

\begin{figure}[hbt]
\centerline{\includegraphics[width=0.7\textwidth]{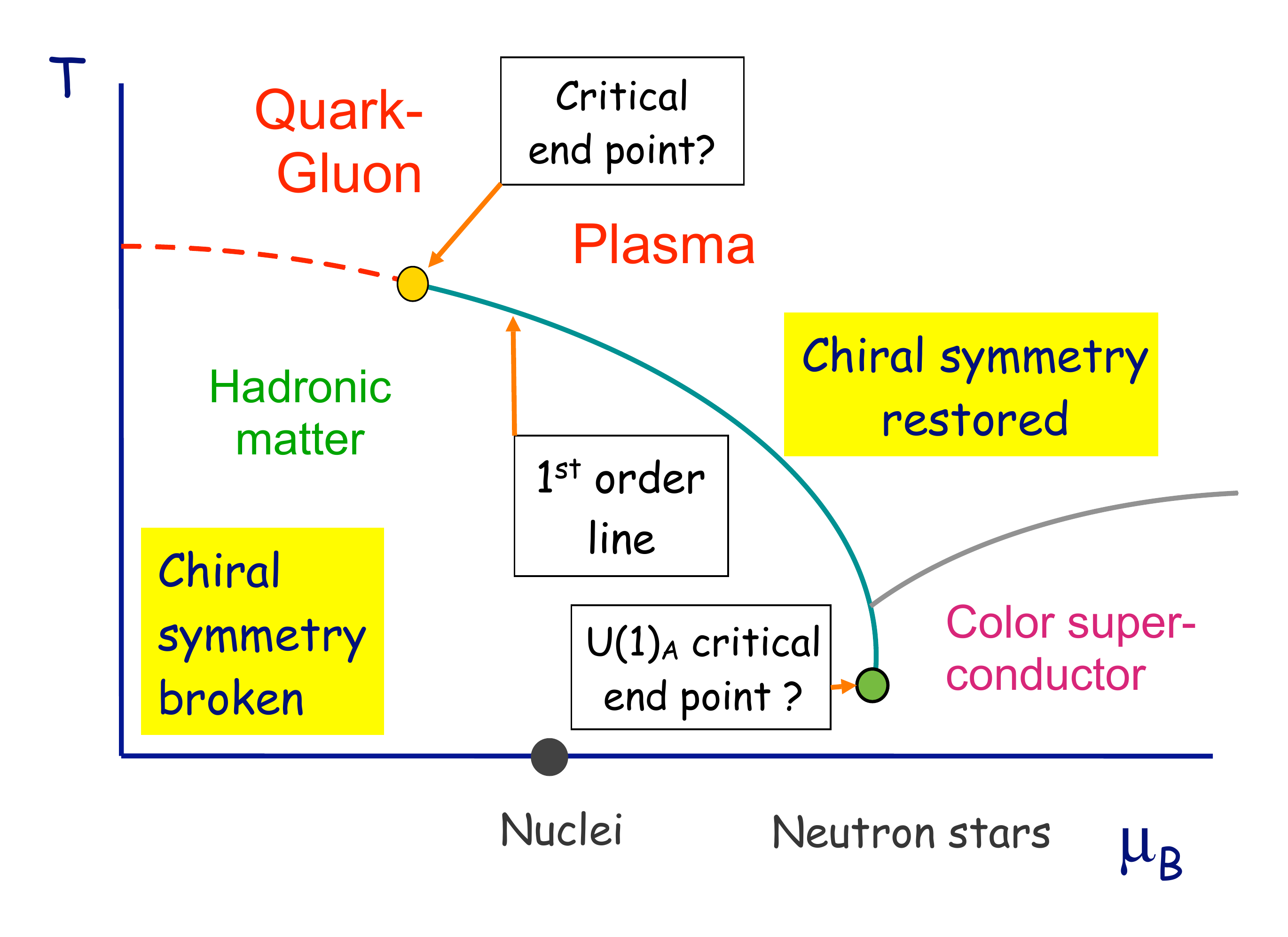}}
\caption{Schematic phase diagram of QCD with two possible critical end-points of the first-order transition line separating the hadronic matter from quark matter.} 
\label{fig:1}
\end{figure}

The strange quark, with  mass $m_s \approx 104\pm 30$ MeV \cite{Amsler:2008zz} comparable to the confinement scale of QCD, influences the QCD phase diagram (see Fig.~\ref{fig:1}) in many subtle, but important ways:

\begin{itemize}
\setlength{\itemsep}{0pt}
\item Lattice simulations have shown that the location of the QCD critical point is extremely sensitive to the value of the strange quark mass $m_s$.
\item By enabling the 'tHooft coupling in the effective chiral Lagrangian, strange quarks may be responsible for the existence of a second critical point on the phase boundary between hadronic and quark matter \cite{Hatsuda:2006ps} in the high-baryon density, low-temperature region.
\item The structure of neutron stars, especially the possibility of a quark matter core, is strongly dependent on $m_s$.
\item The structure of the colour superconducting phase of quark matter is critically sensitive to the value of $m_s$; the colour-flavour locked phase \cite{Alford:1998mk}, in particular, could not exist without strangeness.
\end{itemize}

It thus comes as no surprise that strange quarks -- as well as the heavier flavours, charm and beauty -- are effective probes of hadronic matter. Two aspects are most important in this respect:
\begin{itemize}
\setlength{\itemsep}{0pt}
\item Strange quarks are not easily produced below the critical temperature $T_c$; they thus serve as indicators of quark deconfinement. Charm and bottom quarks are hard to produce thermally at any temperature accessible in laboratory experiments.
\item Quark flavour is conserved under the strong interactions, implying that $s$, $c$, and $b$ quarks most likely survive up to the final state once they have been produced in a reaction.
\end{itemize}
Charm and bottom are also {\em hard probes}, i.e.\ their production can be reliably calculated with methods of perturbative QCD, making them potentially valuable probes for final-state interactions.

The experimental investigation of the structure and properties of hot QCD matter, as it is produced in relativistic heavy ion collisions, requires a theoretical understanding of how experimental observables are related the physical properties of the matter from which they originate. Of particular interest are observables that have a direct connection to those properties which distinguish a quark-gluon plasma from ordinary nuclear or hadronic matter: the deconfinement of quarks and the partial restoration of chiral symmetry. Quark deconfinement at high temperature is closely linked to the phenomenon of colour screening: In the presence of thermally excited gluons and quarks, the colour flux-tube which binds quarks together in the vacuum never forms. A genuine order parameter for (de-)confinement only exists in a world without light quarks, when a vanishing fundamental SU(3) Polyakov loop indicates the absence of free quarks. On the other hand, a genuine order parameter for chiral symmetry restoration, the scalar quark condensate, only exists in a world with massless quarks. At physical quark masses, both quantities make a smooth, but steep transition across the same temperature region between $0.9\,T_c$ and $1.1\,T_c$. 

\section{Quark Liberation and Strangeness Enhancement}

The chemical equilibration of strangeness in the abundances of hadrons produced in a nuclear reaction requires three conditions to be satisfied:
\begin{enumerate}
\item the liberation and thermal equilibration of gluonic degrees of freedom;
\item the reduction in the effective strange quark mass due to the partial restoration of chiral symmetry; and
\item the local deconfinement of quarks causing an increase of the flavour equilibration volume beyond typical hadronic scales.
\end{enumerate}
In a seminal article \cite{Rafelski:1982ii} published in 1982, J.~Rafelski pointed out that in a baryon-rich scenario ``we almost always have more $\bar{s}$ than $\bar{u}$ or $\bar{d}$ quarks.  When quark matter reassembles into hadrons, some of the numerous $\bar{s}$ quarks may, instead of being bound into kaons, form multiply strange antibaryons, such as $\overline{\Lambda}, \overline{\Xi}, \overline{\Omega}$.'' This prediction rested on the assumption that strange quarks are produced in chemical equilibrium with $u$ and $d$ quarks, when quark matter is formed. The question whether this would be so was answered in another article \cite{Rafelski:1982pu}, coauthored by Rafelski and myself, in which we showed that reactions among thermal gluons are the essential mechanism by which strange quark phase space can be equilibrated during the lifetime of the fireball created in a nuclear collision at relativistic energies. This work, followed by a systematic study of the manifestations of strange quark equilibration in hadron yields \cite{Koch:1986ud}, laid the foundation for major experimental programs at the BNL-AGS, CERN-SPS, and now at RHIC. It also has been one of the foundations of the {\em Strangeness in Quark Matter} conference series. 

\begin{figure}[hbt]
\centerline{\includegraphics[width=0.8\textwidth]{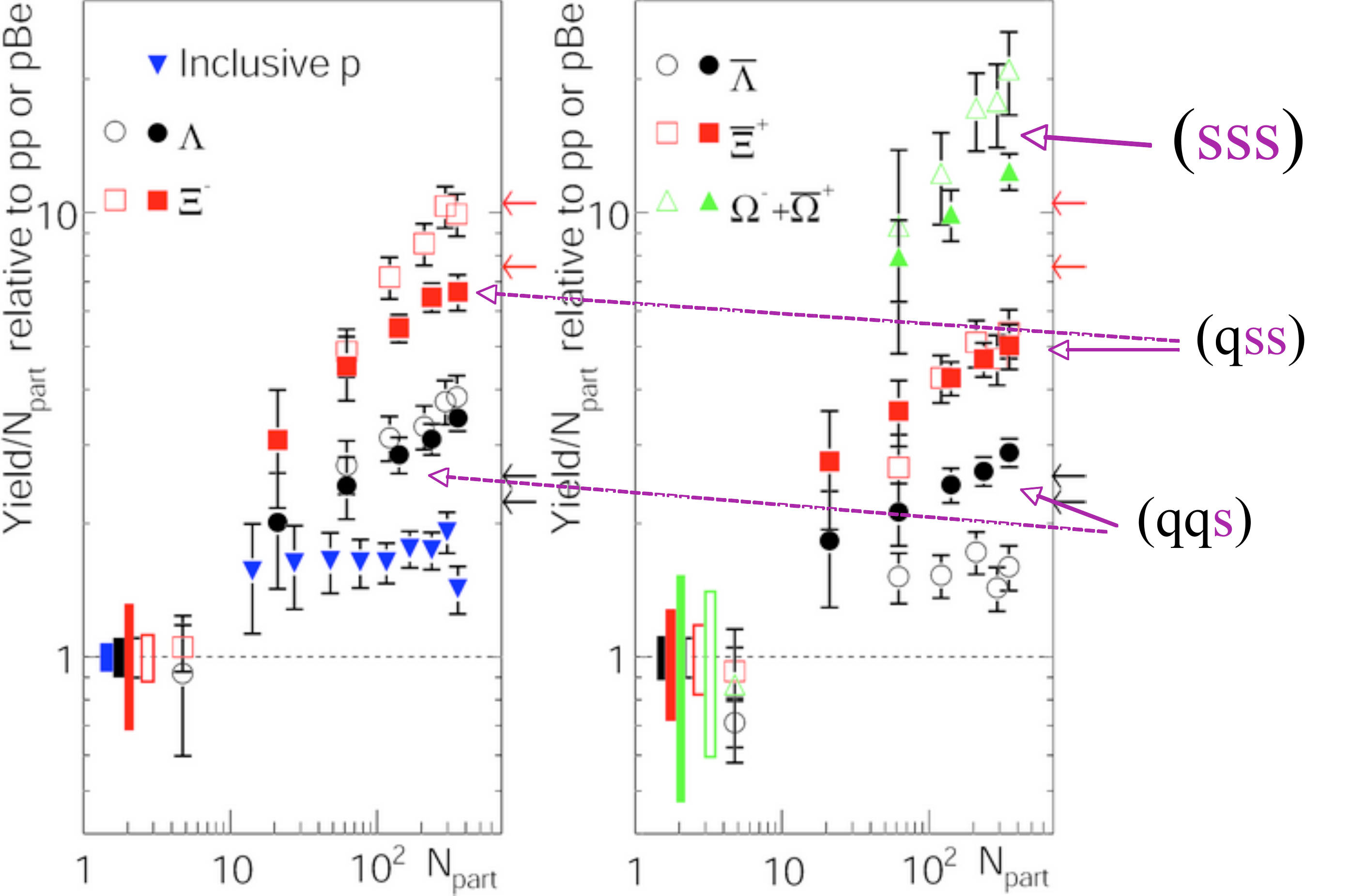}}
 \caption{Centrality dependence of the strange antibaryon enhancement observed in Au+Au at RHIC (from \cite{Takahashi:2008yt}). } 
 \label{fig:2}
\end{figure}

The original predictions \cite{Koch:1986ud} made over 20 years ago have been verified to a remarkable extent. Already at the SPS, the strangeness content of the emitted baryons has been found to increase dramatically with the size of the produced fireball, culminating in an enhancement of the $Omega$ and $\overline{\Omega}$ yield in Pb+Pb collisions by a factor of about 20 \cite{Antinori:2006ij}. The same effect is obesrved at RHIC (see Fig.~\ref{fig:2}), although the enhancement is not quite as strong (only up to a factor 10 for $\Omega(\overline{\Omega})$ in central Au+Au collisions, due to the higher p+p baseline) \cite{Estienne:2004di}. The fraction of strange hadrons grows with increasing beam energy, culminating in perfect equilibration among all light flavours ($u$, $d$, and $s$) in collisions of Au nuclei at the highest RHIC energies
\cite{Speltz:2005qy,Takahashi:2008yt}. The transverse momentum spectra of multi-strange baryons are found to exhibit a steeper slope than expected by virtue of their large masses, indicating little flow effect and suggesting an early freeze-out of the multi-strange baryons from the fireball. The equilibration of strangeness is showing all the signs indicative of abundant production of strange quarks via excitation of gluonic degrees of freedom in hot QCD matter, in combination with the deconfinement of quarks beyond hadronic distance scales.

\begin{figure}[tb]
\centerline{\includegraphics[width=0.7\textwidth]{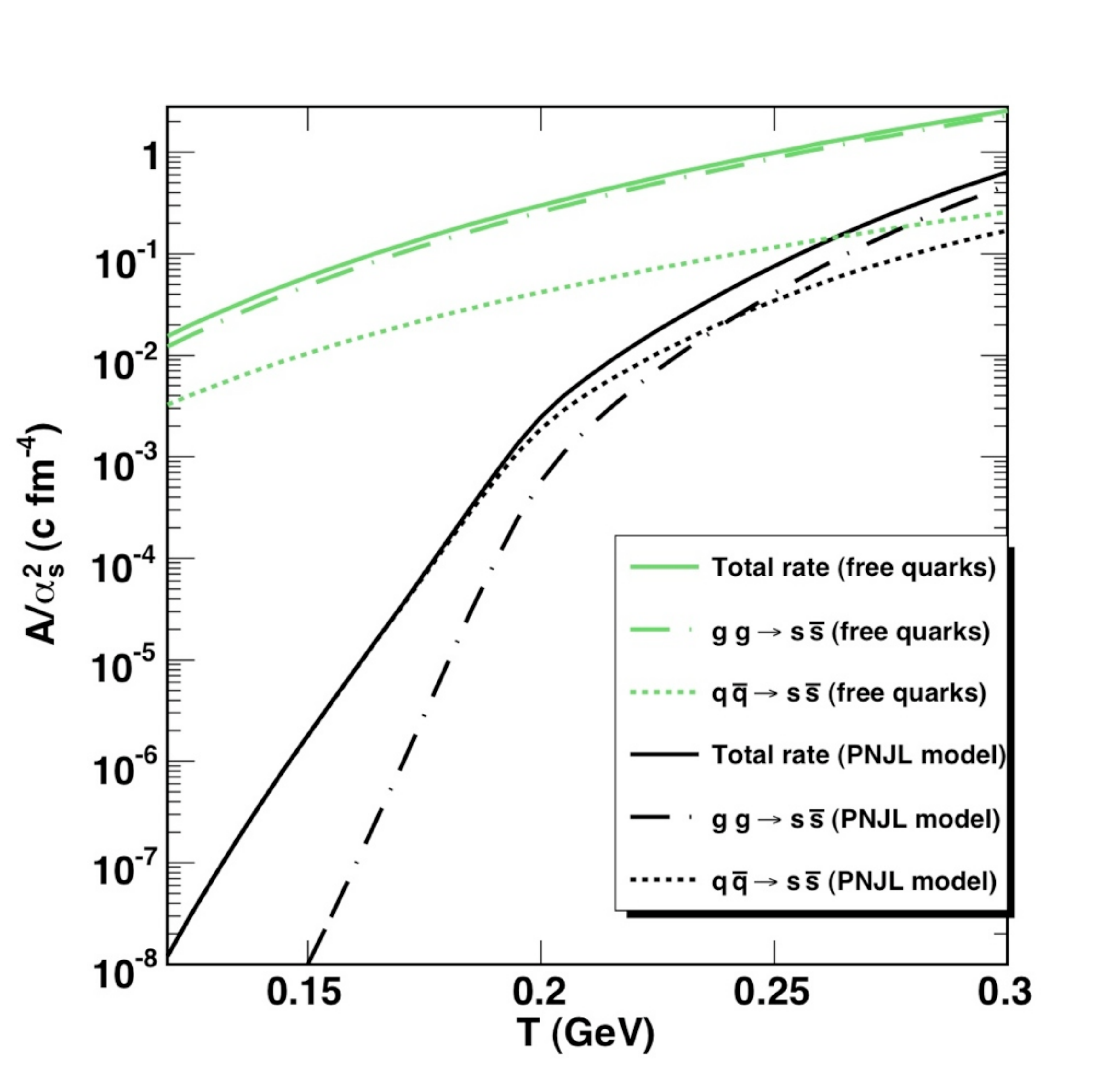}}
 \caption{Temperature dependence of the thermal rates of strange quark-pair production with (green) and without (black) Polyakov loop (from \cite{Tsai:2008je}).}
 \label{fig:3}
\end{figure}

In order to investigate the role of deconfinement on the strange quark production rate, it is useful to calculate strange quark pair production in the three-flavour PNJL model \cite{Fu:2007xc,Ciminale:2007sr}. The PNJL model provides an effective model of QCD near the critical temperature by treating the dynamics of the Polyakov loop (controlling quark confinement) and the chiral quark condensate (controlling chiral symmetry breaking) on an equal footing \cite{Fukushima:2003fw}. While the PNJL model has been mostly applied only to the quark sector, it can also be used to describe the suppression of gluons in the vicinity of $T_c$. Neglecting interactions apart from the Polyakov loop, the spectrum of thermally excited gluons can be written as
\begin{equation}
f_g(k) = \frac{1}{8} \sum_{n=1}^\infty \langle {\rm tr}\, L_A^n \rangle e^{-n|k|/T} 
\quad \stackrel{\langle L_A\rangle \to 1}{\longrightarrow} \quad
\left( e^{|k|/T} - 1\right)^{-1} .
\end{equation}
The expectation value of the adjoint Polyakov loop, $\langle L_A\rangle$, falls faster with decreasing temperature than the expectation of the fundamental Polyakov loop, which controls the suppression of the quark spectrum. In fact, the two expectation values scale exponentially with the eigenvalues of the Casimir operator \cite{Gupta:2007ax}: 
\begin{equation}
\langle {\rm tr}\, L_A \rangle \approx \langle {\rm tr}\, L_F \rangle^{C_A/C_F} = \langle {\rm tr}\, L_F \rangle^{9/4} .
\end{equation}
A graduate student at Duke, Hung-Ming Tsai, has calculated the Polyakov-loop modified quark and gluon spectra and used them to obtain the of thermal rates of strange quark production in the quark-gluon plasma \cite{Tsai:2008je}, which include the effects of the onset of confinement and chiral symmetry breaking in the vicinity of $T_c$. His results, shown in Fig.~\ref{fig:3}, confirm the expectation that the gluonic reaction $gg\to s\bar{s}$ rapidly shuts off as the temperature of the quark-gluon plasma approaches $T_c$ from above. The gluonic rate is found to fall below the light quark rate at $T\approx 230$ MeV. Both rates are significantly smaller than expected in na\"ive perturbation theory. A rough estimate suggests that strangeness equilibration requires initial conditions in the range $T > 300$ MeV. It would be interesting to study the impact of such a scenario in a complete simulation of bulk evolution, e.g.\ multi-fluid hydrodynamics.

\section{Beyond Equilibrium}

The quality of the data from RHIC on yields and spectra of identified hadrons, including resonances, has reached the point where it makes sense to ask questions going beyond the assumption of a stationary equilibrium at freeze-out. This may be our best route to learning more about the hadronization process itself. For example, we may ask: What is the interplay between the dynamical process of hadronization and the inevitable chemical reactions among hadrons in the final state? 

It is well known that particle number changing reactions incrementally cease as the hadron gas expands and cools, forcing the quark fugacities $\gamma_s$ and $\gamma_q$ to become larger than unity, even if $\gamma_s = \gamma_q =1$ at the moment of hadronization.  The assumption of chemical equilibrium inherent in one-fluid hydrodynamics underpredicts protons and antiprotons by factor of three at the kinetic freeze-out temperature $T_f^{(kin)}  \approx 100$ MeV. An round-about way of treating this dynamics is to use an off-equilibrium equation of state for the expanding hadron gas \cite{Hirano:2002ds}. A better approach to the problem is to treat the hadronic gas microscopically in the framework of Boltzmann transport, which allows one to propagate any initial phase-space distribution of hadrons created at hadronization up to the kinetic freeze-out \cite{Hirano:2005wx,Nonaka:2006yn}.

Well tested theoretical tools (the UrQMD or JAM cascades) exist to study the implications for final-state flavour observables of any conceivable hadronization scenario.  Systematic comparisons of different hadronization scenarios with the SPS and RHIC (and soon, LHC) data would be interesting, because it could tell us what the data actually tell us about the hadronization mechanism.

One of the surprising findings of the early RHIC experiments was an unexpected enhancement in the yield of baryons, in comparison with mesons, at transverse momenta in the few GeV/c range. We now understand that this puzzle, as well as some other unexplained experimental observations, can be resolved by the hypothesis that hadrons in an intermediate momentum range are preferentially produced by the recombination of quarks from a thermal quark-gluon plasma \cite{Fries:2003vb,Fries:2003kq,Greco:2003xt,Greco:2003mm,Hwa:2004ng}. The most remarkable prediction of this coalescence mechanism is that the collective flow anisotropy (the elliptic flow $v_2$) in semi-peripheral collisions should satisfy a {\em scaling law}, which relates the elliptic flow of a hadron to the collective flow of quarks via the number of valence quarks of the hadron. The scaling law has been experimentally shown to hold for all measured hadrons, including multi-strange hadrons. What is surprising is that the scaling law appears to hold even in the kinematic region of small transverse energies, where there is no good theoretical justification for it --- unless, that is, the quark-gluon plasma breaks up quite suddenly, with few final-state interactions occurring among the partons.

One mechanism that could lead to such a sudden break-up is a large spike in the bulk viscosity of hot QCD matter right at $T_c$. Lattice simulations \cite{Meyer:2007dy} as well as general considerations \cite{Karsch:2007jc} suggest the presence of such a spike associated with the strong violation of conformal invariance of QCD in the transition domain. A large bulk viscosity $\zeta$ can induce thermodynamic instabilities \cite{Fries:2008ts} (if the effective pressure $P_{\rm eff}=P-\zeta\nabla\cdot v$ becomes negative) or hydrodynamic instabilities, if the solution becomes unstable against density fluctuations \cite{Torrieri:2008ip}. Such scenarios, which so far have only been investigated in the framework of the Bjorken model of linear, boost-invariant expansion of the fireball, clearly need further examination.

\section{Probing Beyond the Hadronic Horizon}

Hadrons containing solely strange quarks (at the valence quark level), such as the $\Omega$ and $\phi$, or heavy quarks, such as charmonium, allow us to probe beyond the hadronization hypersurface. The STAR collaboration has studied the emission of $\Omega$-hyperons and $\phi$-mesons from Au+Au collisions in remarkable detail \cite{Abelev:2007rw}. The analysis shows that the $\phi$ behaves like a meson with respect to the valence quark scaling law for the elliptic flow predicted by quark recombination, and the $\Omega/\phi$ ratio as a function of transverse momentum follows nicely the recombination model predictions \cite{Hwa:2004ir}. The RHIC data also suggest that the collective transverse flow velocity of multi-strange hadrons is lower than that of hadrons composed of up and down quarks, indicative of an early decoupling of those particles after hadronization. Multi-strange hadrons ($\phi, \Xi, \Omega$) may thus be utilized as probes of the kinetic and kinematic characteristics of the hot matter at the moment of its conversion to hadrons.

Charm and bottom quarks have masses that are sufficiently large so that their production is essentially independent of the medium. Their final-state transport and hadronization, on the other hand is strongly affected by the medium. Evidence for the influence of the medium are the suppression factor $R_{AA}$ and elliptic flow $v_2$ of non-photonic electrons observed at RHIC \cite{Adler:2005xv,Abelev:2006db,Sakai:2007zzb}. Bound states of two heavy quarks dissolve into the medium due to colour screening and thermal ionization above $T_c$ \cite{Matsui:1986dk} --- and then form by recombination when the temperature drops below $T_c$  \cite{Thews:2000rj,Andronic:2003zv}.

A comprehensive framework for the description of the interactions of heavy quarks in a QCD medium can be formulated in heavy-quark effective theory \cite{Brambilla:2008cx}. If the colour force is screened at sufficiently short distance, the bound states of a heavy quark-antiquark pair, which exist in the vacuum, may be dissolved.  The quantitative question of the precise location of the dissociation point has been widely investigated using lattice or effective potential techniques, with varying conclusions \cite{Asakawa:2003re,Datta:2003ww,Mocsy:2007yj,Mocsy:2007jz}. Only recently has the full power of hard-thermal loop effective theory been applied to study this problem in the high-temperature limit \cite{Burnier:2007qm}. Since at least for the charmonium states the dissociation points lie close to $T_c$, where HTL effective theory is unreliable, it would be useful to extend the formulation to include nonperturbative properties of the medium, such as temperature dependent gluon condensates \cite{Morita:2007hv}. On the experimental side, the low-transverse mass spectrum of $J/\psi$ measured in Au+Au collisions at $\sqrt{s_{\rm NN}} = 200$ GeV is very close to that measured for the $\Omega$ hyperon and, again like the $\Omega$, has a significantly higher slope parameter than that measured at the lower SPS energies \cite{NuXu:Erice08}. If this is not a coincidence, it suggests that most low-$p_T$ charmonia are formed from thermalized charm quarks at or near $T_c$. The increase from SPS to top RHIC energy would then simply signal the increased transverse flow generated during the deconfined phase at RHIC. The formation of $J/\psi$ by statistical recombination of perturbatively produced, but later thermalized charm quarks also provides the most convincing explanation of the curious rapidity dependence of the nuclear suppression factor observed in Au+Au collisions at RHIC \cite{Andronic:2008gm}.

\section{Summary and Outlook}

Strange, charm, and bottom quarks are powerful probes of the properties of the hot QCD matter formed in relativistic heavy ion collisions. The existing data from RHIC can be best explained by the hypothesis that strange and charm quarks are deconfined during the early phase of the reaction, and that most multi-strange hadrons and many heavy quark bound states form near the hadronization boundary, suffering only rare interactions after production. 

The future on the experimental side appears bright, by virtue of the ongoing upgrades of the STAR and PHENIX detectors at RHIC, combined with the RHIC luminosity upgrade, and the startup of the LHC heavy ion program. On the theoretical side, it will be important to identify the model independent medium properties which can be determined with the help of quark flavour probes (susceptibilities, chemical relaxation times, diffusion coefficients, correlation/screening lengths, etc.) and develop robust observables which can be measured with precision.

{\em Acknowledgements:} This work was supported in part by a grant from the U.S.\ Department of Energy (DE-FG02-05ER41367).
\medskip


\end{document}